\documentstyle[preprint,aps]{revtex}
\begin{document}
\title{Grand canonical partition functions for multi level para Fermi systems 
of any order}
\author{S. Chaturvedi, and V. Srinivasan}
\address{School of Physics\\
University of Hyderabad\\
Hyderabad - 500 046 (INDIA)}
\maketitle
\begin{abstract}
A general formula for the grand canonical partition function for a para Fermi 
system of any order and of any number of levels is derived. 
\end{abstract}
\vskip1.5cm
\noindent PACS No: 05.30.-d,03.65 Bz, 05.70 Ce
\newpage
Parastatistics$^{1,3}$ was introduced by Green$^1$ long ago as a
generalisation of Bose and Fermi statistics . Subequently it has found
many interesting applications to the paraquark model$^4$ and to 
parastring models$^5$.Green's generalisation, carried out at the level of 
the algebra of creation and annihilation operators, involves introducing
trilinear relations in place of the  bilinear relations which
characterize Bose and Fermi systems. The Fock space of an M-level
para Bose system  of order $p$, where $p$ is any positive
integer, is characterized by the trilinear relations
\begin{equation}
[a_k,\{a_l,a_m\}] = 0 ~;~ [a_k,\{a^{\dagger}_l,a^{\dagger}_m\}]
=2{\delta}_{kl} a^{\dagger}_m+ 2{\delta}_{km}a^{\dagger}_l ~ ;~
[a_k,\{a^{\dagger}_l,a_m\}] = 2{\delta}_{kl}a_m ~~,
\end{equation}
and the supplementary conditions
\begin{equation} 
a_k a^{\dagger}_l|0> = p\delta_{kl}|0> \,\,\,.
\end{equation}
Here the subscripts $k,l,m$ take on values $1\cdots M$.
Similarly, the trilinear relations
\begin{equation}
[a_k,[a_l,a_m ]] = 0~;~ [a_k,[a^{\dagger}_l,a^{\dagger}_m]]
=2{\delta}_{kl} a^{\dagger}_m - 2{\delta}_{km}{a^{\dagger}_l}~
;~[a_k,[a^{\dagger}_l,a_m]] = 2{\delta}_{kl} a_m ~~,
\end{equation}
together with the supplementary conditions $(2)$ define a
para Fermi systems of order $p$. Bose and Fermi statistics arise
from these as a special case correspnding to $p=1$. A convenient
and a physically appealing representation of para systems is provided by 
the Green decomposition. Here the annihilation (creation) operators $a_i~
(a^{\dagger}_i)$ for a para system of order $p$ are expressed as
sums of annihilation (creation) operators  operators
$a_{i\alpha}~ (a^{\dagger}_{i\alpha})$ which carry an extra
label $\alpha$ taking values $1,\cdots,p$.
\begin{equation}
a_i = \sum_{\alpha=1}^{p} a_{i\alpha}~~~,~~~~ a^{\dagger}_{i} =
\sum_{\alpha=1}^{p} a^{\dagger}_{i\alpha}~~~~;~~~~a_{i\alpha}|0>
= 0 ~~~.
\end{equation}
The operators $a_{i\alpha}$ and ${a^{\dagger}}_{i\alpha}$ obey
commutation relations which are partly bosonic and partly
fermionic. For a para Bose system of order $p$ these anomalous
commutation relations are
\begin{eqnarray}
[a_{i\alpha}, a_{j\alpha}] = 0~~~;~~~[a_{i\alpha},
a^{\dagger}_{j\alpha}] =\delta_{ij} \nonumber \\
\{a_{i\alpha}, a_{j\beta}\} =
\{a_{i\alpha},a^{\dagger}_{j\beta}\} = 0~~ if~ \alpha\neq \beta~~~~.
\end{eqnarray}
For a para Fermi system of order $p$, the corresponding relations
are
\begin{eqnarray}
\{a_{i\alpha}, a_{j\alpha}\} = 0~~~ ;~~~
\{a_{i\alpha},a^{\dagger}_{j\alpha}\} =
\delta_{ij} \nonumber \\  
{ [a_{i\alpha},a_{j\beta}]~~;~~
[a_{i\alpha},a^{\dagger}_{j\beta}] = 0~~ if~
\alpha \neq \beta}~~~.
\end{eqnarray}

Given the relations $(1)$ and $(3)$ and the supplementary condition $(2)$
one can build a Fock space for parbose and para Fermi systems by
repeated applications of the creation operators on the vacuum state.
For the Fock space thus obtained, two natural question arise. 
\begin{itemize}
\item[[1]] What is the dimensionality of the $N$-particle subspace of 
the Fock space thus obtained? \\
\noindent
and, at a finer level,
\item[[2]] How many independent states are there corresponding to a given set 
of occupation $n_1\cdots n_M$.
\end{itemize} 
One needs answers to these questions,in particular, that of the latter,
in order to be able to construct the canonical
and grand canonical partition functions for parasystems. 
Though parastatistics was introduced nearly forty years ago, it is
seems surprising that, until recently, the only results that were available for
a non trivial parasystem pertained to the following special cases:
\begin{itemize} 

\item[(a)] Single level parasystems of arbitrary order$^{6,7}$

\item[(b)] Two level parasystems of arbitrary order$^{8,9}$

\item[(c)] Parasystems of order two consisting of arbitrary number of 
levels$^{10}$. 
\end{itemize}
The task for the general case i.e. for the case of a parasytem of 
arbitarary number of levels and of arbitarary order was completed recently 
by one of us$^{11}$. This work  encompasses not only parastatistics of any 
order  but also all statistics that can be defined on the basis of the 
permutation group including those for which no simple definition in terms 
of the algebra of creation and annihilation operators is possible. This was
achieved by following the approach to parastatistics pioneered
by  Messiah  and Greenberg$^{12}$  and  further  developed  by
Hartle  Stolt and Taylor$^{13}$. In  this approach parastatistics
arises  in  the  quantum mechanical description of an assembly
of $N$-identical  particles with the permutation group $S_N$
playing a central  role in defining various kinds of statistics
including the parastatistics of Green. In this work, it was shown that
the the general structure of the canonical partition function for an ideal 
system corresponding to any quantum statistics based on the permutation group 
is as follows
 
\begin{equation}
Z_N (x_1,\cdots,  x_M)  =  \sum_{\lambda} 
s_\lambda(x_1,\cdots, x_M)\,\,\,\,,
\end{equation}
where $x_i \equiv \exp(-\beta\epsilon_i) ~;~i=1,\cdots,M$ , $\epsilon_i$
denote the energies corresponding to the states $i=1,\cdots,M$ and 
$\lambda \equiv (\lambda_1,\cdots,\lambda_M); \lambda_1\geq\lambda_\geq\cdots
\lambda_M$ denotes a partition of $N$. The functions
$s_\lambda(x_1,\cdots,x_M)$ denote the Schur functions$^{14,15}$ which are
homogeneous symmetric polynomials of degree $N$ in the variables 
$x_i,\cdots,x_M$. Explicitly the Schur functions are given by 
\begin{equation}
s_\lambda(x_1,\cdots,x_M) =  {\det(x_i^{\lambda_j+M-j)}\over  \det 
(x_i^{M-j})} ~~ ; ~~ 1 \le i,j \le M \,\,\,\,.
\end{equation}
An alternative definition of Schur functions in terms of the monomial 
symmetric functions$^{14,15}$ is as follows

\begin{equation}
s_\chi (x_1,\cdots, x_M)= \sum_{\lambda}  K_{\chi\lambda}
m_\lambda (x_1,\cdots, x_M) \,\,\,,
\end{equation}
where $K_{\chi\lambda}$ denote the Kostka numbers$^{14,15}$.

The canonical partition functions corresponding to various statistics 
based on the permutation group are obtained by putting appropriate 
restrictions on the partitions $\lambda$ that occur on the R.H.S. of
$(7)$. Parabose case of order  $p$ arises when we retrict the sum in $(7)$ 
to only  those partitions of  $N$  whose  length  $l(\lambda)$  (the  number
of   the   non-zero $\lambda_i$'s) is less than  equal  to  $p$. Similarly, 
para Fermi case of order  $p$  arises  when  we restrict $\lambda$ in $(7)$ to 
those partitions for which $\lambda_1\leq p$. 

In this letter we shall confine ourselves to para Fermi systems of order $p$. 
The canonical partition function for such a system is given by 

\begin{equation}
Z_N^{PF}(x_1, \cdots, x_M  ;  p)  =  \sum_{\lambda
\atop{\lambda_1 \leq p}} s_\lambda(x_1,\cdots, x_M)\,\,\,\,,
\end{equation}
and hence the grand canonical partition function is given by

\begin{equation}
{\cal{Z}}^{PF}(x_1, \cdots, x_M, \mu  ;  p) = \sum_{N} exp(\mu\beta N) 
Z_N^{PF}(x_1, \cdots, x_M  ;  p)~~~~.
\end{equation}
Using $(10)$ and the fact that the Schur functions are homogeneous
polynomials of degree $N$, $(11)$ can be written as 
\begin{equation}
{\cal{Z}}^{PF}(x_1, \cdots, x_M, \mu  ;  p)= \sum_{N} \sum_{\lambda\atop
{\lambda_1 \leq p}} s_\lambda(X_1,\cdots, X_M)~~~~~, 
\end{equation}
where $X_i \equiv \exp(-\beta(\epsilon_i - \mu))$. The sum on the R.H.S of
$(12)$ is explicitly known$^{14}$ and is given by

\begin{equation}
{\cal{Z}}^{PF}(X_1, \cdots, X_M  ;  p) = 
{\det(X_j^{2M+p+1-i}-X_j^{i})\over  \det 
(X_j^{2M+1-i}-X_j^{i})} ~~ ; ~~ 1 \le i,j \le M ~~~.
\end{equation}

We now consider some special cases

\noindent{\it Case I: p=1, M arbitrary}

This case corresponds to the Fermi statistics. In this case $(13)$ becomes
\begin{equation}
{\cal{Z}}^{F}(X_1, \cdots, X_M) = 
{\det(X_j^{2M+2-i}-X_j^{i})\over  \det 
(X_j^{2M+1-i}-X_j^{i})} ~~ ; ~~ 1 \le i,j \le M \,\,\,\,.
\end{equation}
In $\det(X_j^{2M+2-i}-X_j^{i})$, adding to each row the one that succeeds it 
and using 
\begin{equation}
(X_j^{2M+2-i}-X_j^{i})+(X_j^{2M+1-i}-X_j^{i+1})= 
(1+X_j) (X_j^{2M+1-i}-X_j^{i})~~~,
\end{equation}
\begin{equation}
(X_j^{M+2}-X_j^{M}) = (1+X_j)(X_j^{M+1}-X_j^{M})~~~~,
\end{equation}
one finds that 
\begin{equation}
\det(X_j^{2M+2-i}-X_j^{i}) = \prod_j (1+X_j)~~ \det(X_j^{2M+1-i}-X_j^{i})~~~~,
\end{equation}
and hence
\begin{equation}
{\cal{Z}}^{F}(X_1, \cdots, X_M) = \prod_j (1+X_j)~~~~~.
\end{equation}

\noindent {\it Case II: p $\rightarrow \infty \; , M \;$ arbitrary }

In this limiting case, christened HST statistics in ref $11$, $(13)$ becomes

\begin{equation}
{\cal{Z}}^{HST}(X_1, \cdots, X_M) = 
(-1)^{M}{\det(X_j^{i})\over  \det 
(X_j^{2M+1-i}-X_j^{i})} ~~ ; ~~ 1 \le i,j \le M ~~~~,
\end{equation}
which on using 
\begin{equation}
\det(X_j^{i}) = \prod_{j} {X_j}^j \prod_{i < j}(1-X_i /X_j)~~~,
\end{equation}
\begin{equation}
\det(X_j^{2M+1-i}-X_j^{i}) =\prod_{j} {X_j}^j \prod_{j} (X_j -1)
\prod_{i < j}(1-X_i /X_j)(1-X_i X_j)~~~~,
\end{equation}
yields 

\begin{equation}
{\cal Z}^{HST}(X_1,\cdots,X_M) =
\prod_i{1\over(1-X_i)}\prod_{i<j} {1\over(1-X_i X_j)}\,\,\,\,.
\end{equation}

\noindent{\it Case III: p arbitrary, M =1}

For $M=1$, with $p$ arbitrary, $(13)$ gives 
\begin{equation}
{\cal{Z}}^{PF}(X;p) = 
{(1-X^{p+1})\over  
(1-X)}~~~~.
\end{equation}

\noindent{\it Case IV: p arbitrary, M =2}

In this case, $(13)$ gives the following expression for the grand canonical 
partition function 

\begin{equation}
{\cal{Z}}^{PF}(X_1, X_2  ; p) =\frac{1}{(1-X_1)} \frac{1}{(1-X_2)}\left[
\frac{(1-(X_1 X_2)^{p+2})}{(1-X_1 X_2)} - \frac{({X_1}^{p+2}- {X_2}^{p+2})}
{(X_1 - X_2)}\right]~~~~.
\end{equation}
This result is the same as that given in ref $9$ but differs from that in ref 
$8$. The mistake in ref $8$ appears to be that the results for para Fermi 
systems were obtained from those for para Bose systems by simply putting 
occupancy restrictions without taking into account the fact that the number 
of states corresponding to a given set of occupation numbers in the two cases 
is also different.

The R.H.S. of $(24)$ may be simplified to yield
\begin{eqnarray}
{\cal{Z}}^{PF}(X_1, X_2  ; p) &=& 
\left( \sum_{l=0}^{p} {X_1}^l \right)
\left( \sum_{l=0}^{p} {X_2}^l \right)  
+ X_1 X_2 \left( \sum_{l=0}^{p-2} {X_1}^l \right)
\left( \sum_{l=0}^{p-2} {X_2}^l \right) \nonumber \\
&+& (X_1 X_2)^2
\left( \sum_{l=0}^{p-4} {X_1}^l \right)\left( \sum_{l=0}^{p-4} {X_2}^l \right)
 + \cdots ~~~~~.
\end{eqnarray}

\noindent{\it Case IV: p arbitrary , M =3}

For $M=3$, $(13)$ gives
\begin{equation}
{\cal{Z}}^{PF}(X_1, X_2, X_3  ;  p) = 
{\det(X_j^{p+7-i}-X_j^{i})\over  \det 
(X_j^{2M+1-i}-X_j^{i})} ~~ ; ~~ 1 \le i,j \le 3~~~~. 
\end{equation}
Setting $p=2$ and simplifying the R.H.S. one obtains
\begin{eqnarray}
{\cal{Z}}^{PF}(X_1, X_2, X_3  ;  2)= &1& + (X_1+X_2+X_3) +(X_1+X_2+X_3)^2
\nonumber \\
&+& (X_1 X_2 + X_2 X_3 + X_3 X_1)[(1+X_1)(1+X_2)(1+X_3)-1] \nonumber \\
&+& (X_1 X_2 X_3)^2~~~~.
\end{eqnarray}
Similarly for $p=3$, one obtains
\begin{eqnarray}
{\cal{Z}}^{PF}(X_1, X_2, X_3  ;  3) &=& (1+X_1)(1+X_2)(1+X_3)[1+(X_1+X_2+X_3)
 \nonumber \\
&+&({X_1}^2 + {X_2}^2 +{X_3}^2 +X_1 X_2 + X_2 X_3 + X_3 X_1) + X_1 X_2 X_3
\nonumber \\
&+&({X_1}^2 {X_2}^2 + {X_2}^2 {X_3}^2 +{X_3}^2 {X_1}^2 \nonumber \\ 
&+& X_1 X_2 X_3 (X_1+X_2+X_3)) + (X_1 X_2 X_3)^2]~~~.  
\end{eqnarray}

To conclude, we have given in $(13)$ the general formula for the grand 
canonical partition function for a multi level para Fermi system of arbitrary
order $p$ and have shown how all hitherto known results for para Fermi systems 
corresponding to specific values of $p$ and $M$ arise from it as special 
cases. 
\vskip0.35cm
\noindent{\bf Acknowledgements:} One of us (SC) is grateful to 
Dr. M Gould for discussions.
\newpage
\noindent{\bf References}
\begin{enumerate}
\item H.S. Green, Phys. Rev. {\bf90}, 270 (1953).
\item S.Doplicher, R. Haag and J.E.  Roberts,  Comm.  Math.  Phys. 
      {\bf23}, 199 (1971).
\item Y. Ohnuki and S. Kamefuchi, {\it Quantum  field  theory  and 
      parastatistics} (Springer Verlag, Berlin, 1982); S.N.
Biswas in {\it Statistical Physics}, eds. N. Mukunda, A.K.
Rajagopal  and K.P. Sinha, Proceedings of the symposium on fifty
years  of Bose statistics, I.I.Sc. Bangalore (India).
\item O.W. Greenberg Phys.Rev.Lett. {\bf13}, 598 (1964)
\item F. Ardalan and F.Mansouri Phys.Rev. D{\bf9}, 3341 (1974) ;
Phys.Rev.Lett. {\bf56}, 2456 (1986).
\item F.Mansouri and X.Wu Phys.Lett. B{\bf 203}, 417 (1988)
\item S.N.Biswas and A.Das  Mod.Phys.Lett. A{\bf3},549(1988)
\item A.Bhattacharya, F.Mansouri, C.Vaz and L.C.R.Wijewardhana Phys.Lett
{\bf22}, 384 (1989)
\item S. Meljanac, M.Stoic and D.Svartan, Preprint, Rudjer Boskovic
Institute, Zaagreb
\item P. Suranyi, Phys. Rev. Lett. {\bf65}, 2329 (1990).
\item S.Chaturvedi, Phys.Rev. E {\bf ??}, ??? (1996)
\item A.M.L. Messiah and O.W. Greenberg, Phys.  Rev.  B{\bf  136}, 
      248 (1964); O.W. Greenberg and A.M.L. Messiah,  B{\bf
138}, 1155 (1965).
\item J.B. Hartle and J.R.  Taylor,  Phys.  Rev.  {\bf178},  2043 
      (1969); R.H. Stolt and J.R. Taylor, Phys. Rev. D{\bf1},
2226 (1970); J.B. Hartle, R.H. Stolt and J.R. Taylor, Phys.
Rev.  D{\bf2}, 1759 (1970).
\item I.G. Macdonald, {\it Symmetric  functions  and  Hall  polynomials} 
      (Clarendon, Oxford, 1979).
\item  B.E. Sagan {\it The symmetric group} ( Brooks / Cole , Pacific Grove,
        California, 1991)
\end{enumerate}
\end{document}